\documentclass[a4paper]{article}

\usepackage{interspeech2011}
\ninept

\usepackage{ifxetex}
\ifxetex
  \usepackage{unicode-math}
  \usepackage{fontspec}
\else
  \usepackage[utf8]{inputenc}
  \usepackage{mathptmx}
\fi

\usepackage{lipsum}

\usepackage{siunitx}

\usepackage[inline, shortlabels]{enumitem}

\usepackage{import}
\usepackage{adjustbox}
\usepackage{tikz}
\usetikzlibrary{chains}
\usetikzlibrary{shapes.geometric}

\usepackage{cite, IEEEtrantools}

\usepackage{hyperref}
\hypersetup{colorlinks, linkcolor=black, urlcolor=black, citecolor=black}
\usepackage[all]{hypcap}

\usepackage[acronym, shortcuts, nowarn]{glossaries}
\glsdisablehyper
\newacronym{3d}{3D}{three-dimensional}
\newacronym{api}{API}{application programming interface}
\newacronym{av}{AV}{audiovisual}
\newacronym[\glsshortpluralkey={DOF},\glslongpluralkey={degrees of freedom}]{dof}{DOF}{degree of freedom}
\newacronym{ema}{EMA}{electromagnetic articulography}
\newacronym{fem}{FEM}{finite element modeling}
\newacronym{gui}{GUI}{graphical user interface}
\newacronym{ik}{IK}{inverse kinematics}
\newacronym{mri}{MRI}{magnetic resonance imaging}
\newacronym{nla}{NLA}{non-linear animation}
\newacronym{tts}{TTS}{text-to-speech}
\newacronym{uti}{UTI}{ultrasound tongue imaging}
\makeglossaries
\providecommand{\acsrobust}[1]{\texorpdfstring{\protect\glsentryname{#1}}{\glsentryname{#1}}}

\newcommand{\arti}{{\em Artimate}}
\newcommand{\keywords}{articulatory modeling, 3D animation, electromagnetic articulography, audiovisual speech synthesis}

\usepackage{authoraftertitle}
\title{\arti: an articulatory animation framework\\
  for audiovisual speech synthesis}

\makeatletter
\def\name#1{\gdef\@name{#1\\}}
\makeatother
\name{{\em Ingmar Steiner$^{1,2}$, Slim Ouni$^{1,3}$}}

\address{$^1$LORIA Speech Group, Nancy, France;
  $^2$INRIA Grand Est;
  $^3$Université de Lorraine \\
{\small $\{$\nolinkurl{ingmar.steiner}$|$\nolinkurl{slim.ouni}$\}$\nolinkurl{@loria.fr}}}

\hypersetup{pdfinfo={%
  Title={\MyTitle},
  Author={Ingmar Steiner and Slim Ouni},
  Keywords={\keywords},
  Subject={Workshop on Innovation and Applications in Speech Technology,
    March 9-10, 2012, Dublin, Ireland}
}}

\begin{document}

\bstctlcite{IEEEexample:BSTcontrol}

\maketitle
\begin{abstract}
We present a modular framework for articulatory animation synthesis using speech motion capture data obtained with \ac{ema}.
Adapting a skeletal animation approach, the articulatory motion data is applied to a \ac{3d} model of the vocal tract, creating a portable resource that can be integrated in an \ac{av} speech synthesis platform to provide realistic animation of the tongue and teeth for a virtual character.
The framework also provides an interface to articulatory animation synthesis, as well as an example application to illustrate its use with a \ac{3d} game engine.
We rely on cross-platform, open-source software and open standards to provide a lightweight, accessible, and portable workflow.
\end{abstract}
\glsresetall

\section{Background and Motivation}

This paper presents a framework for creating portable kinematic articulatory models for \ac{av} speech synthesis, driven by actual speech data.
We refer to \ac{av} speech synthesis as the process of generating speech and displaying speech-synchronized animation for the face and articulators (viz.\ the tongue, lips, jaw) of a virtual character.
While the interior of the vocal tract is not always visible during the synthesis of speech or speech-like motion, the moments during which it is, and does not appear as expected, can disrupt an otherwise convincing visual experience.
The necessity of accounting for articulatory animation in realistic \ac{av} speech synthesis is widely acknowledged;
in fact, the MPEG-4 (part 2) standard includes articulatory movements among the ``facial action parameters'' \cite{Pandzic2002MPEG4}.

Nowadays, facial animation and full-body movements of virtual characters are commonly driven by motion data captured from human actors, using vertex and skeletal animation techniques, respectively \cite{Thalmann2004Handbook}.
However, conventional motion capture approaches (which rely on optical tracking) cannot be directly applied to intraoral articulatory movements during speech, since the tongue and teeth are not fully visible.
This practical restriction may account for the fact that many state-of-the-art \ac{av} synthesizers suffer from a lack of realistic animation for the tongue (and to a lesser extent, the teeth).
Some systems use simple rules to animate the articulators, others omit them altogether \cite{Deng2007DataDriven}.

Meanwhile, speech scientists have a number of medical imaging modalities at their disposal to capture hidden articulatory motion during speech, including realtime \ac{mri}, \ac{uti}, and \ac{ema}.
Such techniques are commonly used to visualize the articulatory motion of human speakers.
Indeed, the resulting data has been applied to articulatory animation for \ac{av} speech synthesis \cite{Cohen1993Modeling, Pelachaud1994CompAnim, King2001JVCA, Engwall2003SpeCom};
using motion-capture data to animate such models can lead to significant improvements over rule-based animation \cite{Engwall2009FONETIK}.
However, these synthesizers are generally focused towards clinical applications such as speech therapy or biomechanical simulation.

While the lips can be animated using optical tracking and the teeth and jaw are rigid bodies, the tongue is more complex to model, since its anatomical structure makes it highly flexible and deformable.
With the exception of \cite{Pelachaud1994CompAnim}, the majority of previous work has modeled the tongue based on static shapes (obtained from \ac{mri}) and statistical parametric approaches to deforming them by vertex animation \cite{Engwall2003SpeCom} or computationally expensive \ac{fem} \cite{Stone2001JASA, Dang2004JASA, Vogt2007Efficient}.
Moreover, the articulatory models used are generally specific to the synthesizer software, and cannot easily be separated for reuse in other \ac{av} synthesizers.

In this paper, we present a first glimpse at \arti, a novel framework for \ac{3d} vocal tract animation driven by articulatory data.
This framework is designed to serve as a component (``middleware'') for an \ac{av} speech synthesis platform, providing animation of the tongue and teeth of a computer-generated virtual character, synchronized with the character's facial animation.
It differs from previous approaches to articulatory animation synthesis in that it combines motion-capture data from \ac{ema} with skeletal animation to generate a self-contained, animated model.

Instead of implementing its own low-level processing, \arti\ leverages existing software for modeling and animation, focusing on kinematics for efficiency and open standards for portability.
It provides a lightweight, cross-platform component that is specifically designed to be incorporated as a resource into a given \ac{av} synthesizer.

In this way, we hope to bridge the gap between research prototypes for articulatory animation synthesis and the wide range of potential applications which would benefit from more realistic articulatory animation for virtual characters.

%
%
%

\section{Implementation}

The \arti\ framework has a modular design, which reflects the different aspects of its approach. The main modules are
\begin{enumerate*}[(a)]
\item the model compiler,
\item the synthesizer core, and
\item a demo application which illustrates how \arti\ can be used.
\end{enumerate*}
These modules are described in this section.

The basic design of \arti\ is essentially comparable to that of \cite{Balci2007Xface}, but while the latter provides only facial animation, \arti\ focuses on the animation of the tongue and teeth, and features a platform-independent implementation.

The framework is managed using Apache Maven \cite{maven}, which covers the dependencies, installation, and integration, as well as \ac{api} and user documentation and distribution.

\subsection{Model compiler}

\begin{figure}
\includegraphics[width=\linewidth]{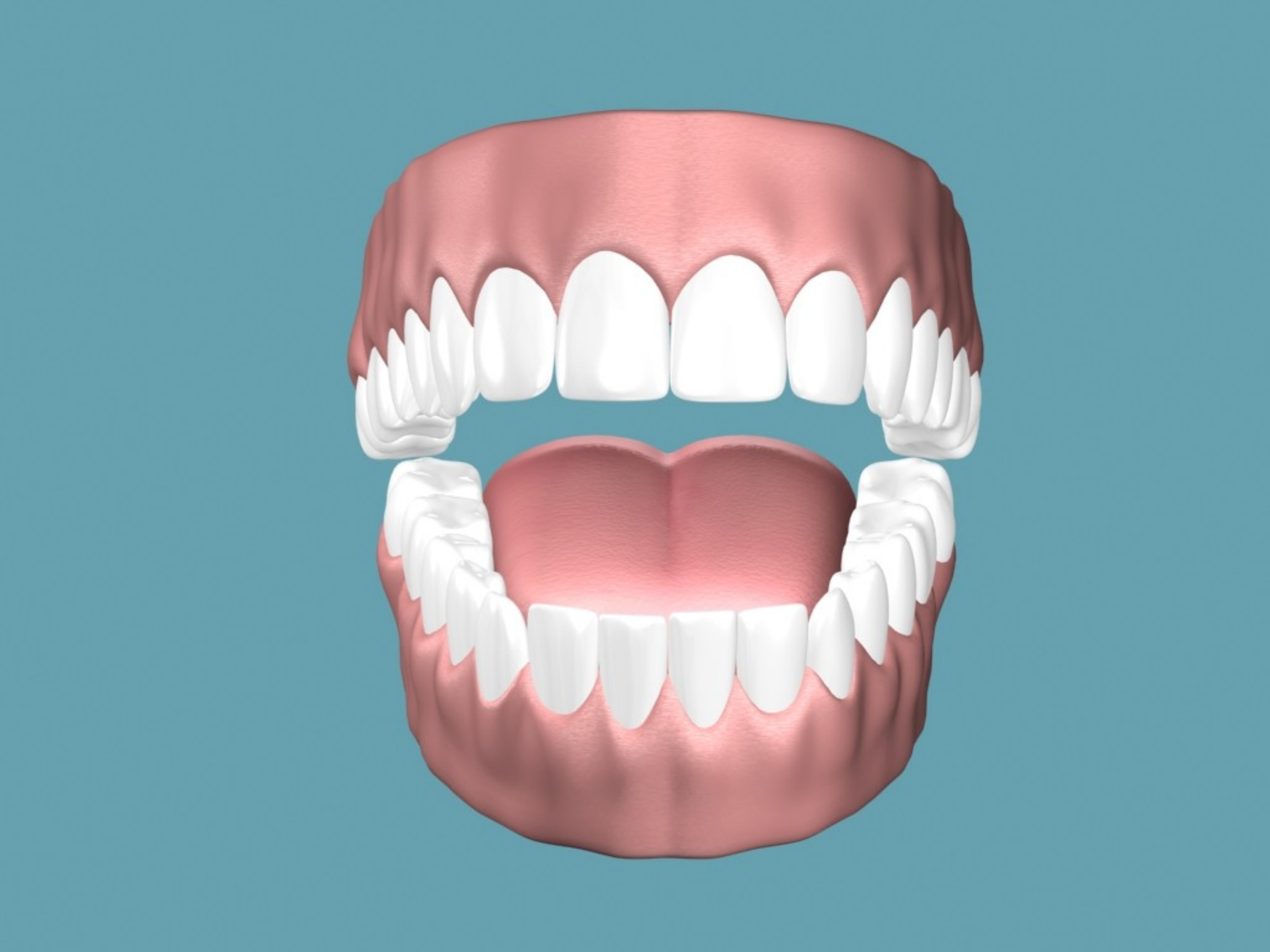}
\caption{Static model of the tongue and teeth, obtained from a stock \ac{3d} model website \cite{turbosquid}.}
\label{fig:stockmodel}
\end{figure}

The model compiler provides a template (``archetype'' in Maven's terminology) which is used to generate an animated \ac{3d} model from appropriate source data.
The user bootstraps the actual compiler from this template, provides speech data in the form of motion capture data obtained from a human speaker using \ac{ema}, and automatically compiles it into an animated model which can be used as a resource by downstream components.

By default, a static \ac{3d} model of the tongue and teeth (\autoref{fig:stockmodel}) is rigged, which was obtained from a stock \ac{3d} model website \cite{turbosquid} under a royalty-free license.
However, the user can provide and configure a custom model to be rigged instead.

The actual processing in this module is performed by automatically generating and running a custom rigging and animation script, which is processed in Blender  \cite{blender}, an open-source, \ac{3d} modeling and animation suite featuring a Python \ac{api}.
The resulting animated model is exported in the open, industry-standard interchange format COLLADA \cite{collada} and bundled as a Java Archive (\texttt{.jar} file), which can be used as a dependency by an \ac{av} synthesizer.

\subsubsection{\acsrobust{ema} data}

\begin{figure}
\def\svgwidth{\linewidth}
{\sffamily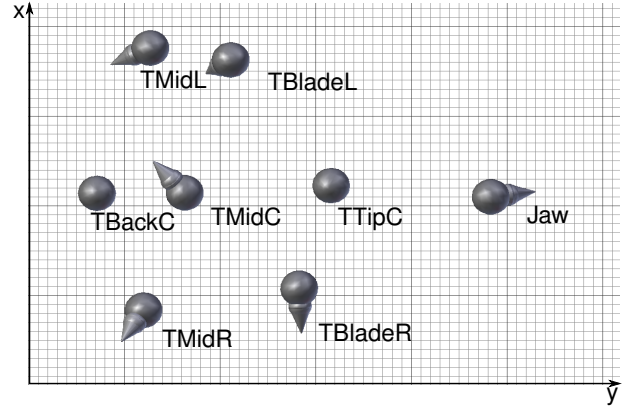}
\caption{\ac{ema} coil layout rendered in the transverse plane (major units in \si{cm}).
The tongue coils are tip center (\textsf{TTipC});
blade left (\textsf{TBladeL}) and right (\textsf{TBladeR});
mid center (\textsf{TMidC}), left (\textsf{TMidL}) and right (\textsf{TMidR});
back center (\textsf{TBackC}).
The absolute coil orientations (rendered as spikes) depend on their attachment in the \ac{ema} recording session.}
\label{fig:coils}
\end{figure}

\arti\ has been developed for use with \ac{ema} data obtained with a Carstens AG500 Articulograph \cite{Hoole20105D}, which provides up to 12 markers (referred to as receiver ``coils'' due to their technical composition) sampled at \SI{200}{\hertz};
each coil has 5~\acp{dof}: the location within the measurement volume (in Cartesian coordinates), and the coil axis orientation (two Euler angles).

Three reference coils are used to normalize for head movement of the speaker, and one is used to track jaw movements, which leaves up to eight coils available to track points (and tangent vectors) on the tongue surface.
The three reference markers are constrained in Blender to lock the others into the scene, irrespective of any external normalization of measured marker positions (e.g., using the Carstens \emph{NormPos} software).

For development, we used a small subset of an existing corpus of \ac{ema} data, featuring seven tongue coils;
three along the mid-sagittal (tongue tip, mid, back center), and two on either side (tongue blade and mid left and right, respectively).
This layout is shown in \autoref{fig:coils}.

\subsubsection{Rigging}

\begin{figure}
\begin{adjustbox}{width=\linewidth}
\begin{tikzpicture}[start chain,
    every node/.style={draw, ellipse, minimum height=7mm, fill={lightgray!50}},
    every path/.style={thick},
    every text node part/.style={font=\sffamily}]
  \node [on chain] {TRoot};
  \node [on chain, join] (tbackc) {TBackC};
  \begin{scope} [start branch=left going above right]
    \node [on chain] (tmidl) {TMidL};
    \draw (tbackc.north) to[bend left] (tmidl.west);
    \node [on chain=going right, join] {TBladeL};
  \end{scope}
  \begin{scope} [start branch=right going below right]
    \node [on chain] (tmidr) {TMidR};
    \draw (tbackc.south) to[bend right] (tmidr.west);
    \node [on chain=going right, join] {TBladeR};
  \end{scope}
  \node [on chain, join] {TMidC};
  \node [on chain, join, xshift=1cm] {TTipC};
\end{tikzpicture}
\end{adjustbox}
\caption{Tongue armature structure for example \ac{ema} coil layout;
labels as in \autoref{fig:coils}, with additional tongue root node (\textsf{TRoot}).}
\label{fig:riggraph}
\end{figure}
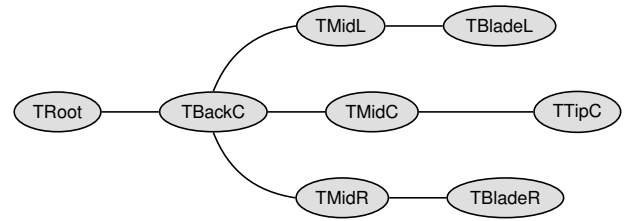

\begin{figure}
\includegraphics[width=\linewidth]{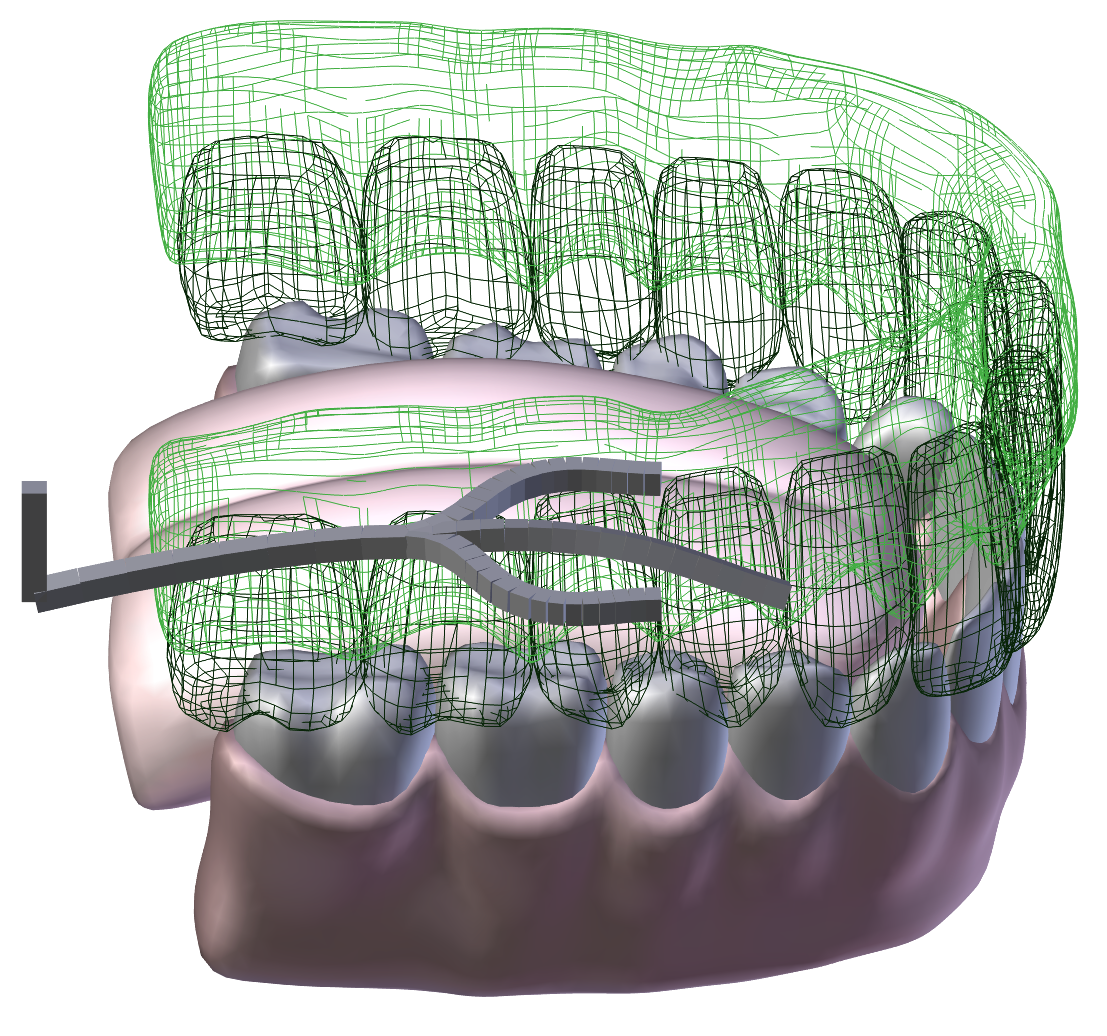}
\caption{Rigged articulatory model with tongue in bind pose.
The maxilla and upper teeth are displayed as wireframe meshes (in light and dark green, respectively), revealing the tongue, mandible, and lower teeth.
The tongue armature is superimposed in gray, posed to meet the \acs{ik} targets (not shown).}
\label{fig:tonguerig}
\end{figure}

The static \ac{3d} model of the tongue and teeth \cite{turbosquid} is rigged with a pseudo-skeleton, or ``armature'' in Blender's terminology, which controls the deformation of the tongue mesh and jaw movements in a standard skeletal animation paradigm, with deformable bones.
The model is configured with several points approximating the layout of the coils in the \ac{ema} data.
These points act as \emph{seeds} for the automatic rigging and animation targets;
the relative movements of the \ac{ema} coils are transferred to the targets as animation data, without the need for registration.

The tongue armature is then assembled using a simple directed graph (\autoref{fig:riggraph}) encoded by a GraphViz \cite{graphviz} \texttt{.dot} file, which defines the structure of the tongue's armature;
the components' rest position is determined by the animation targets' initial positions (\autoref{fig:tonguerig}).
The vertices of the tongue model mesh are then grouped and automatically assigned a weight map which determines the armature components' influence on the position of each vertex.

The armature is constrained to track the animation targets using \ac{ik} \cite{itasc}, while maintaining its volume during stretching and compression.
Additional constraints can be added to the animation targets to smooth coil jitter and counter any measurement errors in the original \ac{ema} data;
otherwise, any such errors will translate directly into noticeably unrealistic movements of the tongue model.

\begin{figure*}
\includegraphics[width=.3\linewidth]{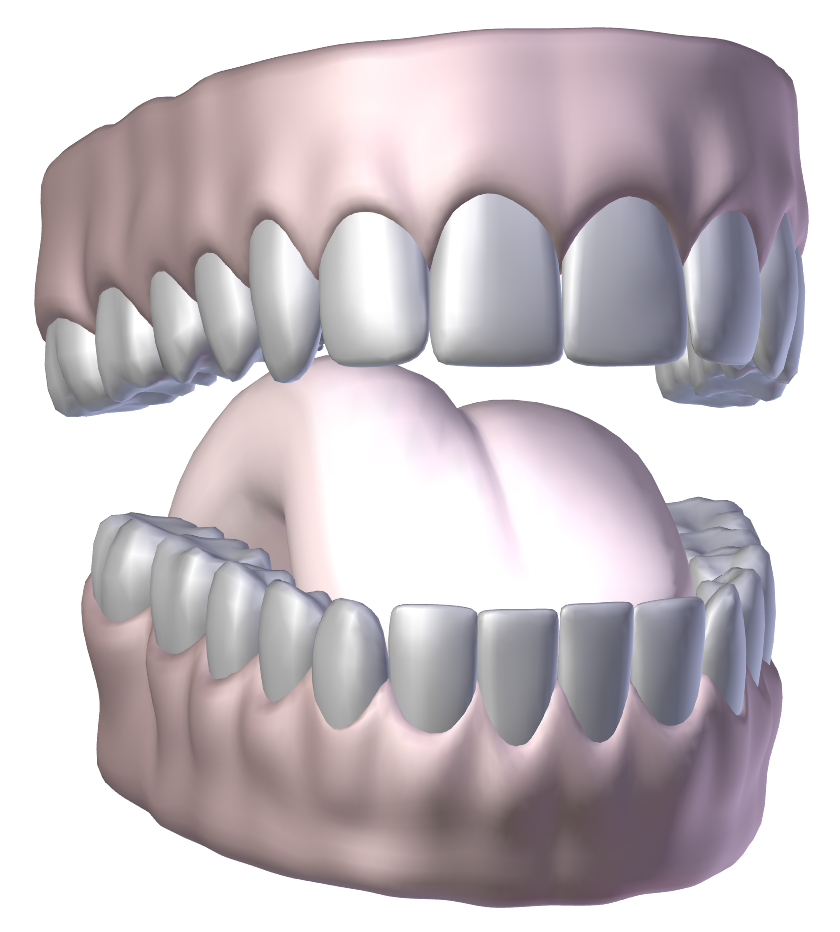}
\hfill
\includegraphics[width=.3\linewidth]{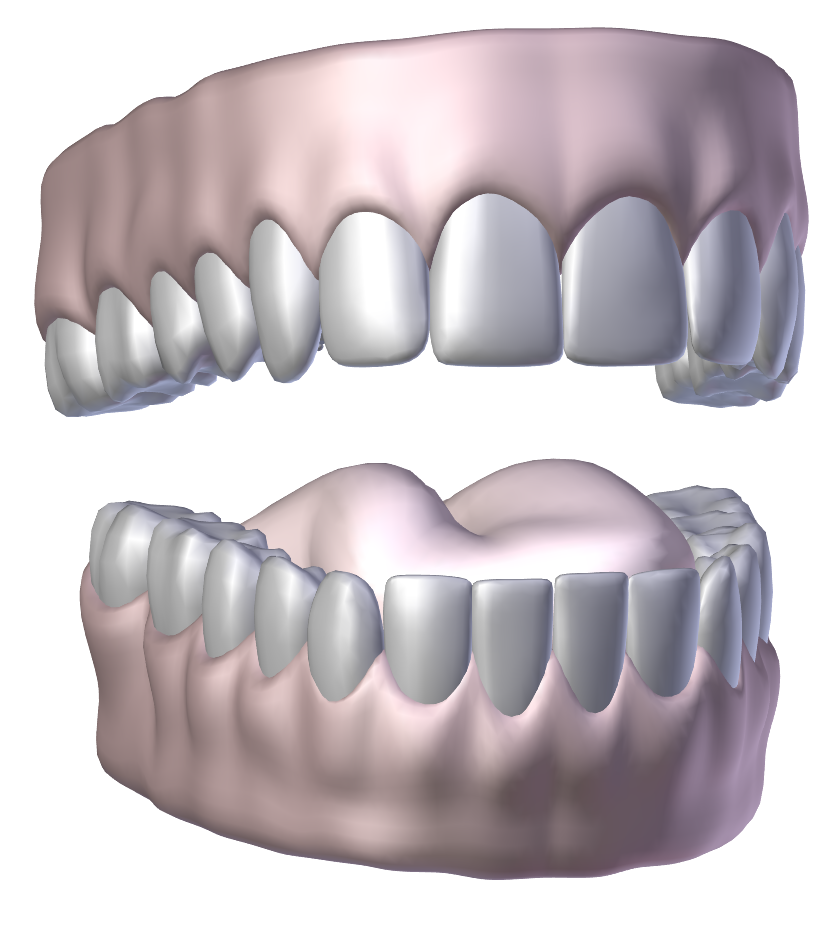}
\hfill
\includegraphics[width=.3\linewidth]{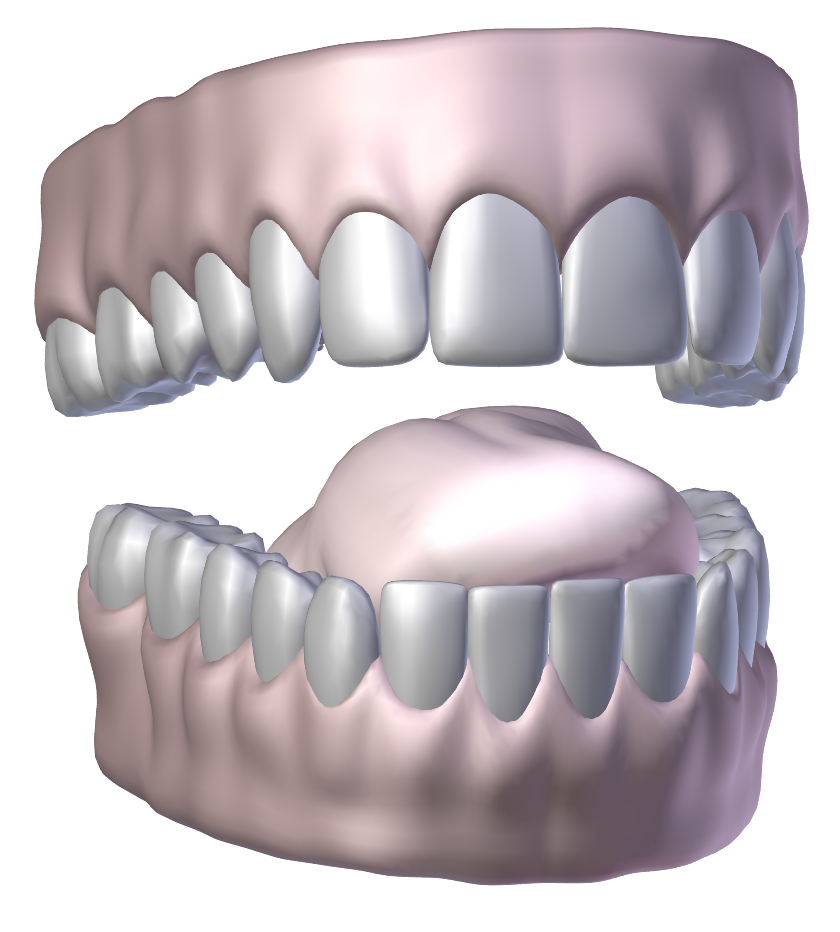}
\caption{The animated tongue model, posed according to several \ac{ema} data frames:
a bunched configuration (left), a grooved tongue blade (center), and during an apical gesture (right).
The asymmetry of the \ac{ema} data has been preserved.
Note that the jaw opening has been exaggerated to improve visibility of the tongue surface.}
\label{fig:animated}
\end{figure*}

Finally, the result is exported as a self-contained, animated \ac{3d} model, which includes the articulatory movements of the human speaker, but is independent of the original \ac{ema} data. Several poses of the tongue model, taken directly from the animation data, are rendered in \autoref{fig:animated}.

\subsubsection{Speech synthesis resources}

The animation is currently generated in a single timeline, represented by the consecutive acquisition sweeps of the \ac{ema} data.
If a phonetic annotation based on the acoustic segments spoken is available, corresponding temporal markers will be created in Blender, and it would be possible to restructure the animation as corresponding ``actions'' in a \ac{nla} paradigm.
However, to maintain portability with external software, the linear animation timeline is used in the COLLADA model.
Nevertheless, the segmentation file is bundled into the generated resource, so that the animation can be split downstream for \ac{nla} synthesis.

The acoustic signals from the \ac{ema} recording session can be included as well, in case the audio is required externally.

\subsubsection{Validation}

In addition to generating the animated model in COLLADA format, the animated model is also saved as a \texttt{.blend} file for interactive exploration and debugging in Blender.
Moreover, the positions of the \ac{ema} coils, \ac{ik} targets, and vertices on the tongue mesh can be dumped to \texttt{.pos} files compatible with the Carstens \ac{ema} format, which can be used to externally visualize and evaluate the generated animation.

This permits validation of the generated articulatory trajectories by direct comparison with the source \ac{ema} data.

\subsection{Core library}

The resource generated by the model compiler includes the animation data derived from \ac{ema}, and optionally, segmentation and/or audio.
This can be directly included into an \ac{av} synthesizer for articulatory animation of virtual characters.
However, in most cases, unless a very low-level, direct control of the articulatory model is desired, it will be preferable to wrap the articulatory animation in a separate library, which exposes a public \ac{api} for articulatory animation synthesis and handles the direct control of the model internally.
This is the purpose of \arti's core library module.

The core library, which is implemented in Java, serves as the interface between the external \ac{av} synthesizer and the articulatory model.
Using a \ac{3d} engine wrapper, it handles loading the articulatory model and provides a lightweight, unit-selection synthesizer dedicated to articulatory animation synthesis.
Animation units are selected from the database of available animation units provided by the animated model's segmentation, using an extensible cost function based on the requested segment durations and smoothness of concatenation.

This core library is simple enough to be extended or ported as needed, depending on external requirements.

\subsection{Demo application}

To test the practicality of reusing the \arti-compiled resource and core library, a demo application was developed, which also serves to illustrate the integration of the framework into an independent platform.
This demo application is implemented in Java and consists of a simple \acl{gui}, which displays an interactive \ac{3d} view of the animated model.

The demo application makes use of Ardor3D \cite{ardor3d}, one of a family of open-source, Java 3D game engines, which features mature COLLADA support.
The core library is used to select animation units from the model, which are rendered by Ardor3D's internal animation system.

In addition to directly requesting specific units from the animation database, the demo application can also generate speech-synchronized animation from text via the multilingual \ac{tts} platform MARY \cite{Schroeder2009BC}.

\section{Conclusions}

We have presented a modular framework for building and deploying a kinematic model for articulatory animation in \ac{av} speech synthesis.
The animation is driven by motion capture in the form of \ac{ema} data and uses a skeletal animation approach with a deformable armature for the tongue.
Models compiled in this manner can be reused as a self-contained resource by external applications, either with the core library as an interface, or directly accessing the model's animation data.

The \arti\ framework will become publicly available under an open-source license in the first half of 2012, hosted at \url{http://artimate.gforge.inria.fr/}.

Future work includes extending model compiler support to other \ac{ema} data formats, such as those produced by alternative processing software (e.g., TAPADM \cite{tapadm}) or acquired with other Carstens or NDI device models.

Furthermore, we will investigate using \arti\ to build an articulatory animation model from a much larger corpus, such as the \texttt{mngu0} articulatory database, whose abundant \ac{ema} data is complemented by volumetric \ac{mri} scans that could be used to extract one or more static vocal tract models \cite{Richmond2011IS, Steiner2012JASA}.

\section{Acknowledgements}

This publication was funded by the French National Research Agency (ANR - ViSAC - Project N. ANR-08-JCJC-0080-01).

\eightpt
\bibliographystyle{IEEEtran}
\bibliography{ref}

\begin{thebibliography}{10}
\providecommand{\url}[1]{#1}
\csname url@samestyle\endcsname
\providecommand{\newblock}{\relax}
\providecommand{\bibinfo}[2]{#2}
\providecommand{\BIBentrySTDinterwordspacing}{\spaceskip=0pt\relax}
\providecommand{\BIBentryALTinterwordstretchfactor}{4}
\providecommand{\BIBentryALTinterwordspacing}{\spaceskip=\fontdimen2\font plus
\BIBentryALTinterwordstretchfactor\fontdimen3\font minus
  \fontdimen4\font\relax}
\providecommand{\BIBforeignlanguage}[2]{{%
\expandafter\ifx\csname l@#1\endcsname\relax
\typeout{** WARNING: IEEEtran.bst: No hyphenation pattern has been}%
\typeout{** loaded for the language `#1'. Using the pattern for}%
\typeout{** the default language instead.}%
\else
\language=\csname l@#1\endcsname
\fi
#2}}
\providecommand{\BIBdecl}{\relax}
\BIBdecl

\bibitem{Pandzic2002MPEG4}
I.~S. Pandzic and R.~Forchheimer, Eds., \emph{{MPEG}-4 Facial Animation: The
  Standard, Implementation and Applications}.\hskip 1em plus 0.5em minus
  0.4em\relax Wiley, 2002.

\bibitem{Thalmann2004Handbook}
N.~Magnenat-Thalmann and D.~Thalmann, Eds., \emph{Handbook of Virtual
  Humans}.\hskip 1em plus 0.5em minus 0.4em\relax Wiley, 2004.

\bibitem{Deng2007DataDriven}
Z.~Deng and U.~Neumann, Eds., \emph{Data-Driven {3D} Facial Animation}.\hskip
  1em plus 0.5em minus 0.4em\relax Springer, 2007.

\bibitem{Cohen1993Modeling}
M.~M. Cohen and D.~W. Massaro, ``Modeling coarticulation in synthetic visual
  speech,'' in \emph{Models and Techniques in Computer Animation},
  N.~Magnenat-Thalmann and D.~Thalmann, Eds.\hskip 1em plus 0.5em minus
  0.4em\relax Springer, 1993, pp. 139--156.

\bibitem{Pelachaud1994CompAnim}
C.~Pelachaud, C.~van Overveld, and C.~Seah, ``Modeling and animating the human
  tongue during speech production,'' in \emph{Proc. Computer Animation},
  Geneva, Switzerland, May 1994, pp. 40--49.

\bibitem{King2001JVCA}
S.~A. King and R.~E. Parent, ``A {3D} parametric tongue model for animated
  speech,'' \emph{Journal of Visualization and Computer Animation}, vol.~12,
  no.~3, pp. 107--115, Sep. 2001.

\bibitem{Engwall2003SpeCom}
O.~Engwall, ``Combining {MRI}, {EMA} {\&} {EPG} measurements in a
  three-dimensional tongue model,'' \emph{Speech Communication}, vol.~41, no.
  2-3, pp. 303--329, Oct. 2003.

\bibitem{Engwall2009FONETIK}
O.~Engwall and P.~Wik, ``Real vs.\ rule-generated tongue movements as an
  audio-visual speech perception support,'' in \emph{Proc. FONETIK}, Stockholm,
  Sweden, May 2009, pp. 30--35.

\bibitem{Stone2001JASA}
M.~Stone, E.~P. Davis, A.~S. Douglas, M.~NessAiver, R.~Gullapalli, W.~S.
  Levine, and A.~Lundberg, ``Modeling the motion of the internal tongue from
  tagged cine-{MRI} images,'' \emph{Journal of the Acoustical Society of
  America}, vol. 109, no.~6, pp. 2974--2982, Jun. 2001.

\bibitem{Dang2004JASA}
J.~Dang and K.~Honda, ``Construction and control of a physiological
  articulatory model,'' \emph{Journal of the Acoustical Society of America},
  vol. 115, no.~2, p. 853–870, Feb. 2004.

\bibitem{Vogt2007Efficient}
F.~Vogt, J.~E. Lloyd, S.~Buchaillard, P.~Perrier, M.~Chabanas, Y.~Payan, and
  S.~S. Fels, ``Efficient {3D} finite element modeling of a muscle-activated
  tongue,'' in \emph{Biomedical Simulation}, ser. Lecture Notes in Computer
  Science, M.~Harders and G.~Székely, Eds.\hskip 1em plus 0.5em minus
  0.4em\relax Springer, 2007, vol. 4072, pp. 19--28.

\bibitem{Balci2007Xface}
K.~Balcı, E.~Not, M.~Zancanaro, and F.~Pianesi, ``Xface open source project
  and {SMIL}-agent scripting language for creating and animating embodied
  conversational agents,'' in \emph{Proc. {ACM} Multimedia}, Augsburg, Germany,
  Sep. 2007, pp. 1013--1016.

\bibitem{maven}
\BIBentryALTinterwordspacing
{Sonatype, Inc.}, \emph{Maven: The Definitive Guide}.\hskip 1em plus 0.5em
  minus 0.4em\relax O'Reilly, 2008. [Online]. Available:
  \url{http://www.sonatype.com/Books/Maven-The-Complete-Reference}
\BIBentrySTDinterwordspacing

\bibitem{turbosquid}
\BIBentryALTinterwordspacing
Bitmapworld, ``Free gums 3d model.'' [Online]. Available:
  \url{http://www.turbosquid.com/FullPreview/Index.cfm/ID/230484}
\BIBentrySTDinterwordspacing

\bibitem{blender}
\BIBentryALTinterwordspacing
Blender. [Online]. Available: \url{http://blender.org/}
\BIBentrySTDinterwordspacing

\bibitem{collada}
\BIBentryALTinterwordspacing
M.~Barnes and E.~L. Finch, \emph{{COLLADA} -- Digital Asset Schema Release
  1.5.0}, Khronos Group, Apr. 2008. [Online]. Available:
  \url{http://collada.org/}
\BIBentrySTDinterwordspacing

\bibitem{Hoole20105D}
P.~Hoole and A.~Zierdt, ``Five-dimensional articulography,'' in \emph{Speech
  Motor Control: New developments in basic and applied research}, B.~Maassen
  and P.~van Lieshout, Eds.\hskip 1em plus 0.5em minus 0.4em\relax Oxford
  University Press, 2010, ch.~20, pp. 331--349.

\bibitem{graphviz}
\BIBentryALTinterwordspacing
E.~R. Gansner and S.~C. North, ``An open graph visualization system and its
  applications to software engineering,'' \emph{Software: Practice and
  Experience}, vol.~30, no.~11, pp. 1203--1233, 2000. [Online]. Available:
  \url{http://graphviz.org/}
\BIBentrySTDinterwordspacing

\bibitem{itasc}
J.~De~Schutter, T.~De~Laet, J.~Rutgeerts, W.~Decré, R.~Smits, E.~Aertbeliën,
  K.~Claes, and H.~Bruyninckx, ``Constraint-based task specification and
  estimation for sensor-based robot systems in the presence of geometric
  uncertainty,'' \emph{International Journal of Robotics Research}, vol.~26,
  no.~5, pp. 433--455, 2007.

\bibitem{ardor3d}
\BIBentryALTinterwordspacing
{Ardor3D}. [Online]. Available: \url{http://ardor3d.com/}
\BIBentrySTDinterwordspacing

\bibitem{Schroeder2009BC}
\BIBentryALTinterwordspacing
M.~Schröder, S.~Pammi, and O.~Türk, ``Multilingual {MARY} {TTS} participation
  in the {B}lizzard {C}hallenge 2009,'' in \emph{Proc. Blizzard Challenge},
  Edinburgh, UK, Sep. 2009. [Online]. Available: \url{http://mary.dfki.de/}
\BIBentrySTDinterwordspacing

\bibitem{tapadm}
\BIBentryALTinterwordspacing
A.~Zierdt. Three-dimensional {A}rtikulographic {P}osition and {A}lign
  {D}etermination with {MATLAB}. [Online]. Available:
  \url{http://wiki.ag500.net/TAPADM}
\BIBentrySTDinterwordspacing

\bibitem{Richmond2011IS}
K.~Richmond, P.~Hoole, and S.~King, ``Announcing the electromagnetic
  articulography (day 1) subset of the mngu0 articulatory corpus,'' in
  \emph{Proc. Interspeech}, Aug. 2011, pp. 1505--1508.

\bibitem{Steiner2012JASA}
I.~Steiner, K.~Richmond, I.~Marshall, and C.~D. Gray, ``The magnetic resonance
  imaging subset of the mngu0 articulatory corpus,'' \emph{Journal of the
  Acoustical Society of America}, vol. 131, no.~2, pp. 106--111, Feb. 2012.

\end{thebibliography}

\end{document}